\documentclass[aps,floats,amssymb,amsmath,prb,twocolumn,superscriptaddress]{revtex4}
\usepackage{calc}
\usepackage{psfrag}
\usepackage{graphicx}
\usepackage{color}
\newcommand{\ttilde}{\tilde{t}}

\begin{document}
\title{$d$-wave  superconductivity on the honeycomb bilayer}

\author{J. Vu\v ci\v cevi\'c}
\affiliation{Scientific Computing Laboratory, Institute of Physics
Belgrade, University of Belgrade, Pregrevica 118, 11080 Belgrade,
Serbia.}

\author{M. O. Goerbig}
\affiliation{Laboratoire de Physique des Solides, Université Paris-Sud, CNRS UMR 8502, F-91405 Orsay Cedex, France.}

\author{M. V. Milovanovi\'c}
\affiliation{Scientific Computing Laboratory, Institute of Physics
Belgrade, University of Belgrade, Pregrevica 118, 11080 Belgrade,
Serbia.}

\begin{abstract}
We introduce a microscopic model on the honeycomb bilayer, which in the small-momentum limit captures the usual (quadratic dispersion in kinetic term)
description of bilayer graphene.
 In the limit of strong interlayer hopping it reduces to an effective honeycomb monolayer model with also third neighbor hopping. We study
interaction effects in this effective model focusing on possible superconducting instabilities. We find $d_{x^2-y^2}$ superconductivity in
the strong coupling limit of an effective $tJ$-model-like description that gradually transforms into $d + i d$ time-reversal symmetry breaking
superconductivity at weak couplings.
In this limit the small momentum
order parameter expansion is $(k_x + i k_y)^2$ [or $(k_x - i
k_y)^2$]  in both valleys of the effective low-energy
description. The relevance of our model and investigation for
the physics  of bilayer graphene is also discussed.
\end{abstract}

\maketitle

\section{Introduction}

Interaction effects are expected to be
important for the physics of bilayer graphene and may cause
a formation of correlated many-body phases.\cite{Castro,Kotov}
This needs to be contrasted to intrinsic monolayer graphene  in which a vanishing
density of states at the Dirac points suppresses the influence of electronic correlations.\cite{Kotov,Goerbig}
Recent experiments on suspended bilayer graphene,
\cite{Martin, Weitz,Freitag,Lau} which is free of substrate effects,
reveal a gapped state at and around the charge neutrality
point. The state may be of topological origin\cite{Qi} due to the observed\cite{Martin,Freitag}
conductance of the order of $e^2/h$
and may exhibit
an anomalous quantum Hall effect, i.e. a quantum Hall effect at zero
magnetic field. In the most recent experiment on high mobility samples
from Ref.~\onlinecite{Lau}, a completely insulating behavior was found.

From the theory point of view, several proposals were given
\cite{Min,
Nandkishore,Zhang,Vafek,FZhang,Jung,Lemonik,ZhangMacD,Vafek_three,Kharitonov,Scherer}
for the existence of gapped (and gapless) phases at the charge
neutrality point, including those that break the time-reversal
symmetry. Most of them are based on the particle-hole (excitonic)
binding which is the most natural assumption in the understanding
of a gapped phase at the charge neutrality point.  These
theories assume a quadratic dispersion of the electrons in the
low-energy effective description \cite{McFa}, and direct hopping between two
sublattices in different layers that leads to the linear
dispersion (``triagonal warping") is neglected. This assumption is justified
if the chemical potential is not exactly situated at the charge-neutrality point.

To explore additional possibilities for gapped
phases in the presence of a finite chemical potential, we
discuss here superconducting instabilities, especially with an eye on the
possibility of topological (fully gapped) superconductivity on
the honeycomb bilayer.  Bilayer graphene may be potentially also viewed
as a strongly-correlated system with a possibility to support a layered
antiferromagnetic state,\cite{FZhang,Jung} similar
to the Mott physics of high $T_c$ superconductors. The existence of a layered antiferromagnetic state
is supported by the most recent experiment with high quality samples, \cite{Lau}
which feature completely insulating behavior at the charge neutrality point.

There is, so far, no systematic study of
superconducting instabilities in the presence of electron-electron
and electron-phonon interactions on the honeycomb bilayer at finite doping
(see, however, Ref.~\onlinecite{Vafek_two} for fermions in the presence of
weak electron-electron interactions only at zero chemical potential). To address this question,
we study in the present paper a microscopic model
of a single effective honeycomb monolayer with reduced nearest
neighbor hopping and third-neighbor hopping, in addition to inter-site attractive interactions.
The kinetic term of the effective model is obtained by integrating out the ``high-energy''
degrees of freedom from the direct interlayer hopping (i.e. assuming strong interlayer hopping in the honeycomb bilayer), and the inter-site
superexchange interaction originates from the Hubbard on-site repulsion.
This model is to a certain degree biased to antiferromagnetism and
$d$-wave superconductivity, but preserves the usual low-energy description of the
bilayer graphene.\cite{McFa} Moreover, in contrast to the usual low-energy model of
bilayer graphene, the present model accounts for the lattice symmetry of the original
model (the honeycomb bilayer) that may be relevant for the symmetry of the superconducting order parameters.
%in this sense has a potential to
%be relevant, at least for its low energy description.

Our primary interest here is to find the most probable symmetry of a
superconducting instability on the honeycomb bilayer together with an understanding of its nature i.e. whether this instability is topological.
We also aim at an understanding of the change in the superconducting
order parameter and correlations as we go from a monolayer to a
few-layer honeycomb lattice. The mean-field solution of the introduced model yields a
time-reversal symmetry breaking $d+id$-wave superconducting state
at weak coupling, which  continuously transforms into
$d_{x^2-y^2}$-wave with increasing interaction. Near 3/8 and 5/8
filling of the $\pi$-bands, i.e.~near the van-Hove singularity in
the density of states, the Cooper pairing becomes much stronger.
Our conclusion  is that  the $d + i d$ superconducting instability
is the leading superconducting instability of the honeycomb bilayer with strong interlayer hopping at finite doping and the same instability
may be present in the bilayer graphene at finite doping. However,
due to the presumed smallness of coupling constant and order
parameter, as well as strong quantum fluctuations in two
dimensions, it may be difficult to detect this order
experimentally in today's graphene samples.

The remaining part of the paper is organized as follows. In Sec. II we define our effective two-band
model on an effective honeycomb lattice with third-nearest-neighbor hopping. The
model is then, in Sec. III, solved by a Bogoliubov - de Gennes (BdG) transformation for a singlet bond-pairing
order parameter, and we discuss the relevant symmetries. Section IV
presents the phase diagram obtained from a numerical solution of the BdG equations.
In Sec. V, the relevance for the physics of the bilayer graphene is discussed,
and our main conclusions are presented in Sec. VI. Two Appendices
summarize analytically obtained solutions in the weak-coupling BCS limit.

\section{Model}

The honeycomb bilayer lattice consists of two Bernal-stacked honeycomb lattices, each consisting of
two triangular sublattices as illustrated in Fig.~1 such that the unit cell contains four lattice sites.
The Hamiltonian of free electrons on such a lattice is given by
\begin{eqnarray}
H_{0}& = &
         - t \sum_{\vec{j},\sigma} \sum_{\vec{u}}
             \left(   a_{1,\vec{j},\sigma}^{\dagger} b_{1, \vec{j} + \vec{u},\sigma}
               + a_{2,\vec{j},\sigma}^{\dagger} b_{2, \vec{j} -\vec{u},\sigma} + \mbox{H.c}\right) \nonumber \\
&&       - t_{\bot} \sum_{\vec{j},\sigma}
             \left(   a_{1,\vec{j},\sigma}^{\dagger} a_{2, \vec{j},\sigma} + \mbox{H.c}\right) \nonumber \\
&&       - \mu \sum_{i,\vec{j}}
           \left(  a_{i,\vec{j},\sigma}^{\dagger}a_{i,\vec{j},\sigma}
                 + b_{i,\vec{j},\sigma}^{\dagger} b_{i, \vec{j},\sigma}\right). %\nonumber
\label{freeham}
\end{eqnarray}
Here, the index $i = 1,2$ denotes the layer and $\vec{j}$ enumerates primitive cells. The sum runs over
$\vec{u}=\vec{u}_0,\vec{u}_1,\vec{u}_2$, where $\vec{u}_1=a(\frac{3}{2},\frac{\sqrt{3}}{2})$
and $\vec{u}_2=a(\frac{3}{2},-\frac{\sqrt{3}}{2})$ are the primitive vectors of the lattice,
and $\vec{u}_0$=(0, 0) is an auxillary vector for denoting the hopping between sites in the same primitive cell.
The norm of these vectors is $|\vec{u}|=\sqrt{3}a$, in terms of the distance, $a$,  between neighboring
sites in each layer, and $t$  is the associated hopping energy, whereas
$t_{\bot}$  denotes the interlayer hopping energy, between
A sites in two different layers.  The finite chemical potential $\mu$ takes into account doping, either
due to the electric-field effect or to chemically active adatoms.
The operators $a_{i,\vec{n},\sigma}^{\dagger} (a_{i, \vec{n},\sigma})$ represent electron creation (annihilation)
on the sublattice site $A_i$ of the layer $i$ with spin
$\sigma = \uparrow, \downarrow$, and $b_{i,\vec{n},\sigma}^{\dagger}
(b_{i, \vec{n},\sigma})$ those for electrons on the sublattice site $B_i$.
 $\mu$ is the
chemical potential.
We use units such that $\hbar = 1$.

\begin{figure}
\centering
\includegraphics[width=7cm]{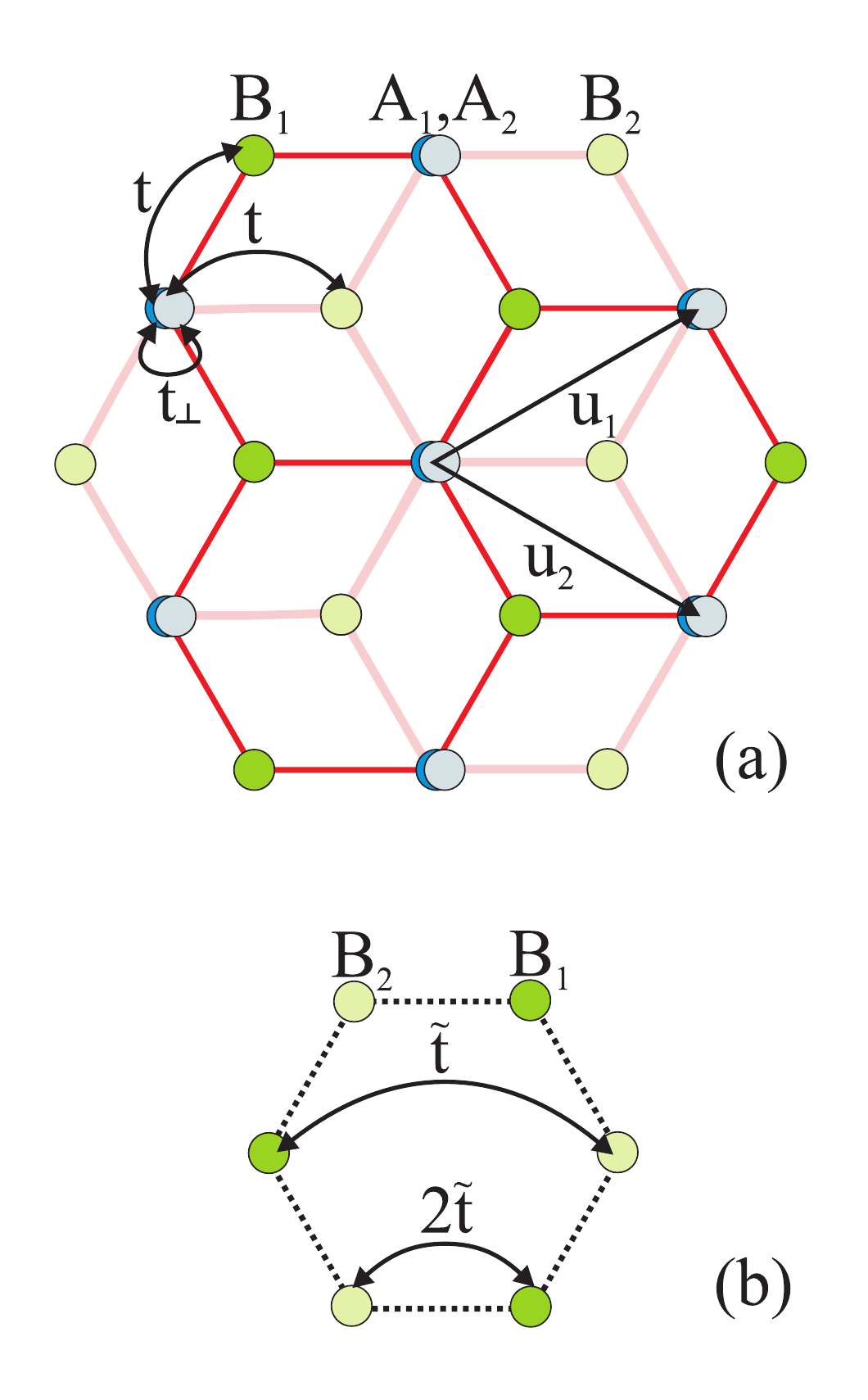}
\caption{(Color online) (a) A view of Bernal stacked honeycomb
lattices 1 and 2 with corresponding sublattice sites A1, B1, and
A2, B2 respectively. (b) The model reduces to a monolayer model
with the third neighbor hopping $\tilde t \equiv t^2 / t_\bot$ and
the nearest neighbor hopping $2\tilde t$ (see the text). }
\label{fig:01}
\end{figure}

By introducing the Fourier
transforms $ a_{i, \vec{k}, \sigma} = \sum_{\vec{j}} a_{i, \vec{j},
\sigma} \exp( i \vec{k} \cdot \vec{j})$ and $ b_{i, \vec{k}, \sigma}
= \sum_{\vec{j}} b_{i, \vec{j}, \sigma} \exp(  i \vec{k} \cdot
\vec{j})$, and diagonalizing the Hamiltonian, one obtains  the
spectrum
\begin{equation}
E^{\pm}_{\alpha} (\vec{k}) = \pm \left[(-1)^{\alpha} \frac{t_{\bot}}{2} +
\sqrt{\frac{t_{\bot}^{2}}{4} + t^2 |\gamma_{\vec{k}}|^{2}}\right]\;, \label{dispersion_gb}
\end{equation}
where $\alpha = 1,2$ and $\pm$ denote 4 different branches of dispersion and
\begin{equation}\label{gamma}
\gamma_{\vec{k}} = \sum_{\vec u} e^{i\vec{k}\cdot\vec{u}} = 1 + e^{i\vec{k}\cdot\vec{u}_1}+e^{i\vec{k}\cdot\vec{u}_2}.
\end{equation}

In their orginal work,\cite{McFa} McCann and Fal'ko showed that the four-band model may be simplified to an effective
two-band model if one considers energies much smaller than $t_{\bot}$.
In momentum space, the Hamiltonian in Eq.~(\ref{freeham}) becomes
\begin{eqnarray}
H_0 &=& \sum_\sigma \int_{BZ} \frac{d^2 \vec{k}}{(2\pi)^2} \\
&&  \left\{ - t \left( \gamma_{\vec{k}} a_{1,\sigma,\vec{k}}^{\dagger} b_{1,\sigma,\vec{k}} \right.\right.
   + \left.          \gamma^*_{\vec{k}} a_{2,\sigma,\vec{k}}^{\dagger} b_{2,\sigma,\vec{k}} + \text{H.c.}\right)\nonumber \\
&&  -  t_\bot \left(a_{1,\sigma,\vec{k}}^{\dagger} a_{2,\sigma,\vec{k}} + \text{H.c.}\right)\nonumber \\
&&   - \mu \left(a_{1,\sigma,\vec{k}}^{\dagger} a_{1,\sigma,\vec{k}}
        +   a_{2,\sigma,\vec{k}}^{\dagger} a_{2,\sigma,\vec{k}} \right. \nonumber \\
&&\left.\left. +   b_{1,\sigma,\vec{k}}^{\dagger}b_{1,\sigma,\vec{k}}
               +   b_{2,\sigma,\vec{k}}^{\dagger} b_{2,\sigma,\vec{k}}\right) \right\}.
\end{eqnarray}
If we introduce the spinor
\begin{equation}
\Psi_{\sigma}(\vec{k}) = (a_{1,\sigma, \vec{k}}, a_{2,\sigma, \vec{k}},
b_{2,\sigma, \vec{k}}, b_{1,\sigma, \vec{k}})^{T},
\end{equation}
the Hamiltonian can be expressed as a $4\times 4$ matrix,
\begin{equation}
H_0(\vec{k}) = \sum_\sigma \Psi^\dagger_{\sigma}(\vec{k})
        \left[\begin{array}{cccc}  - \mu & -t_\bot & 0 & - t \gamma_{\vec{k}} \\
                                    -t_\bot & - \mu & - t \gamma^*_{\vec{k}} & 0\\
                                    0 & - t \gamma_{\vec{k}} & - \mu & 0 \\
                                    -t \gamma^*_{\vec{k}}& 0 & 0& - \mu\\

\end{array} \right] \Psi_{\sigma}(\vec{k}).
\end{equation}
One may further define $2 \times 2$ matrices $ H_{11} = - \mu I +
t_\bot \sigma_x, \, H_{22} = - \mu I, \, H_{12} = - t (\mathrm{Re} \gamma_{\vec{k}} \sigma_x +
\mathrm{Im} \gamma_{\vec{k}} \sigma_y) = H_{21}$, such that the eigenvalue equation can be written in the
following form ($\vec{k}$ indices are implied)
\begin{equation}
\left[\begin{array}{cc}  H_{11} & H_{12} \\
                                      H_{21}  & H_{22} \\

\end{array} \right]
\left[\begin{array}{c}  \Psi_1 \\
                                      \Psi_2 \\

\end{array} \right]= E
\left[\begin{array}{c}  \Psi_1 \\
                                      \Psi_2 \\

\end{array} \right],
\end{equation}
from which we obtain
\begin{equation}
\{H_{22} - H_{21} (H_{11} - E)^{-1} H_{12}\} \Psi_2 = E
\Psi_2.\label{jed}
\end{equation}
If we assume $t_\bot$ to be the largest energy scale
 and consider the low-energy limit ($E \ll t_\bot$),
Eq.~(\ref{jed}) becomes

\begin{equation}\label{ham:eff}
H_{\mathrm{eff}} \Psi_2 \equiv \left[\begin{array}{cc}  - \mu &  \frac{t^2}{t_\bot} \gamma_{\vec{k}}^2 \\
                                       \frac{t^2}{t_\bot} \gamma_{\vec{k}}^{*2}   & - \mu \\

\end{array} \right] \Psi_2 = E \Psi_2,
\end{equation}
with $ \Psi_2 (\vec {k})= (b_{2,\sigma,\vec{k}}, b_{1,\sigma,\vec{k}})^T.$

The two-band model described by the Hamiltonian in Eq. (\ref{ham:eff}), is also valid in the limit \cite{McFa} where
$E\ll t_{\bot}\ll t$. For energies larger than $t_{\bot}$, one needs to take into account the other two bands which overlap in energy
with those considered in Eq. (\ref{ham:eff}).
In the following sections we use the simplified two-band model at even larger energies, up to the van-Hove singularity.
Formally, this amounts to increasing artificially (with respect to the graphene bilayer) the interlayer hopping $t_{\bot}$ such that it becomes the largest energy scale, $t_{\bot}\gg t$. In that limit Eq. (\ref{ham:eff}) becomes the exact description of the honeycomb bilayer for $E, t \ll t_\bot$ and for the wavevectors of the whole Brillouin zone. We will adopt that model in the following.

The Hamiltonian in Eq. (\ref{ham:eff}) corresponds, in real space,
to a single-layer honeycomb lattice with nearest-neighbor and
third-neighbor hoppings. Whereas the effective hopping amplitude
of the latter is given by $t^2/t_{\bot}$, the effective
nearest-neighbor hopping is twice as large.\cite{Bena} This means that due to
the strong interlayer hopping, the complete low-energy physics is
projected onto the B1 and B2 sublattices which themselves form a
hexagonal lattice (see Fig.~\ref{fig:01}).

\begin{figure}
\includegraphics[width = 8cm]{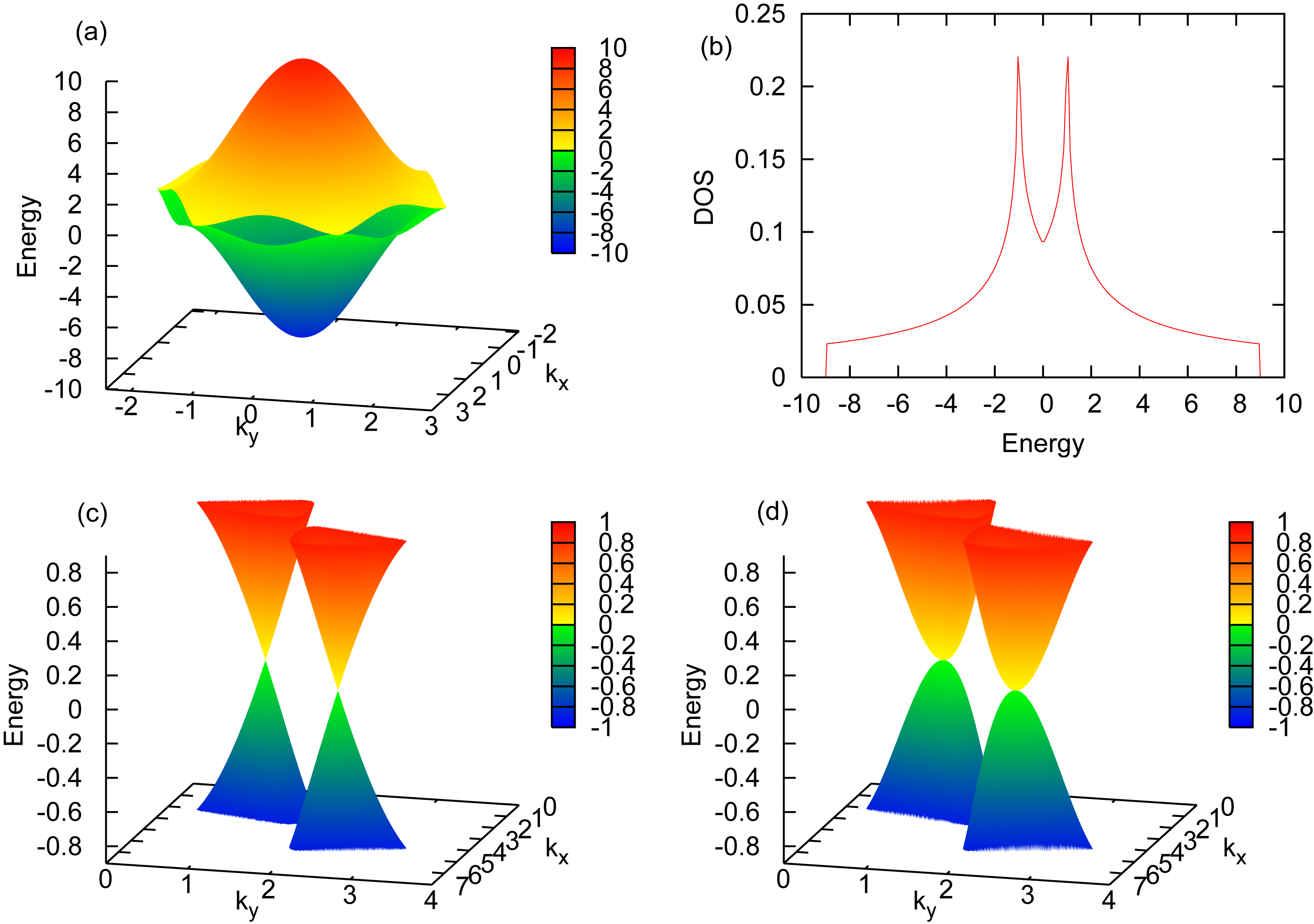}
\caption{(Color online) Non-interacting dispersion (a) and density
of states (b) of the projected monolayer model. Linear dispersion
in the vicinity of the K-points in the graphene monolayer (c) in
comparison to the quadratic dispersion in our model (d). We use
$\tilde t = t^2/t_{\bot}$ for the unit of energy.} \label{fig:02}
\end{figure}

As mentioned above, the
model is equivalent to the graphene bilayer in the small-momentum limit, i.e. for
$t^2/t_\bot |ka|^2 \sim \mu \ll t^2/t_\bot$ and reproduces
correctly the finite density of states (DOS) at $E=0$ of bilayer
graphene [Fig.~\ref{fig:02}].  Finally, the Hamiltonian (\ref{ham:eff}) does not take into
account direct hopping between the B1 and B2 sublattices, which may though easily be accounted
for by adding $-t'\gamma_{\vec{k}}^*$ to the off-diagonal matrix elements, where $t'\simeq 0.3$ eV
is the associated hopping amplitude. This term yields the so-called trigonal warping close to the
charge-neutrality point, which consists of a splitting of the parabolic band-contact point into
four linear Dirac points.\cite{McFa} However, these Dirac points are present only at very low energies,
for chemical potentials $|\mu|$ in the meV range, such that the parabolic-band approximation
becomes valid even at low dopings. Since we are interested, here, in moderate doping, we neglect
this additional term and use the effective band model (\ref{ham:eff})  in the following sections.

Since we consider the effective hopping $t^2/t_{\bot}$ to be small and if there is a
significant on-site repulsion $U$, spin-singlet bonds between B1
and B2 sites are expected to form due to superexchange processes.
Therefore, we apply the $t-J$ model but relax the requirement of
the model that double occupation of sites is excluded. We justify this by our primary aim: to find the most probable symmetry of the superconducting instability. As we will
be working in the mean-field approximation, we just assume an
effective nearest-neighbor attractive interaction between
electrons on B1 and B2 sublattices, and in doing this we favor
spin-singlet bond formation. The spin-singlet formation directly follows from the mean-field approach to the $t-J$ model \cite{Black}.
If the attractive interaction is not too
strong, it can be simply added to  Hamiltonian (\ref{ham:eff}), with the help
of the term

\begin{equation}
H_{I} = - J \sum_{\vec{j},\vec{u}} \sum_{\sigma}
b_{1,\vec{j},\sigma}^{\dagger} b_{1, \vec{j},\sigma} b_{2,\vec{j}+\vec{u},-\sigma}^{\dagger} b_{2, \vec{j}+\vec{u},-\sigma},
\label{interaction}
\end{equation}
where $J > 0$.
Now we apply the BCS ansatz by introducing the superconducting order parameter
as a 3 component complex vector
$$\bold{\Delta}\equiv(\Delta_{\vec{u}_0},\Delta_{\vec{u}_1},\Delta_{\vec{u}_2})$$
where the components are defined by
\begin{equation}
\Delta_{\vec{u}} = \frac{1}{\sqrt{2}}\langle b_{1,\vec{j},\uparrow} b_{2, \vec{j}
+ \vec{u},\downarrow} - b_{1,\vec{j},\downarrow} b_{2, \vec{j}
+ \vec{u},\uparrow} \rangle,
\end{equation}
and correspond to the spin-singlet pairing amplitudes
of three inequivalent pairs of nearest neighbors. The interaction part $H_I$
in the mean-field approximation becomes
\begin{eqnarray}
H_{BCS}& =& \sqrt{2}  J \sum_{\vec{j}, \vec{u}}
\Delta_{\vec{u}} \left( b_{1,\vec{u}, \uparrow}^{\dagger} b_{2,
\vec{j} + \vec{u}, \downarrow}^{\dagger} - b_{1,\vec{j},
\downarrow}^{\dagger} b_{2,
\vec{j} + \vec{u}, \uparrow}^{\dagger} \right) + \mbox{H.c.}  \nonumber \\
&& + 2 N \sum_{\vec{u}} J
|\Delta_{\vec{u}}|^{2},
\end{eqnarray}
where $N$ is the number of unit cells.

%%%%%%%%%%%%%%%%%%%%%%%%%%%%%%%%%%%%%%%%%%%%%%%%%%%%%%%%%%%%%%%%%%%%%%%%%%%%%%%%

\section{Bogoliubov - de Gennes analysis and pairing symmetries}

\begin{figure}
\centering
\caption{ Different pairing instabilities in real space: (a)
s-wave, (b) $d_{x^2-y^2}$ wave, (c) $d_{xy}$ wave, and (d)
$d_{x^2-y^2} + i d_{xy}$ time reversal breaking $d$-wave.}
\label{fig:03}
\end{figure}

The complete BCS Hamiltonian in momentum
space is given by
\begin{eqnarray}
&& H = - \frac{t^2}{t_\bot} \sum_{\vec{k},\sigma} \left(\gamma_{\vec{k}}^2 b_{2,
\vec{k} \sigma}^\dagger b_{1, \vec{k} \sigma} + \text{h.c.}\right) \nonumber \\
&& + \sqrt{2}J \sum_{\vec{k}} \left[ \sum_{\vec{u}}
\Delta_{\vec{u}} e^{i \vec{k} \cdot \vec{u}} \left( b_{2, \vec{k}
\uparrow}^\dagger b_{1, -\vec{k} \downarrow}^\dagger - b_{2, \vec{k}
\downarrow}^\dagger b_{1, -\vec{k} \uparrow}^\dagger \right) + \mbox{H.c.}
\right]\nonumber \\
&&- \mu \sum_{\vec{k},\sigma} \left (b_{1, \vec{k} \sigma}^\dagger b_{1,
\vec{k} \sigma} + b_{2, \vec{k} \sigma}^\dagger b_{2, \vec{k}
\sigma}\right). \label{BCS_HAM}
\end{eqnarray}
Similar to the case of the honeycomb monolayer,\cite{Black} we can make our description much more transparent if
we apply the following transformation that diagonalizes the kinetic
part of the above Hamiltonian,
\begin{equation}
\left[ \begin{array}{c} b_{2, \vec{k} \sigma}\\b_{1, \vec{k}
\sigma}\end{array} \right] = \frac{1}{\sqrt{2}} \left[
\begin{array}{c} d_{\vec{k} \sigma} + c_{\vec{k} \sigma}\\
e^{-i 2 \varphi_{\vec{k}}} (d_{\vec{k} \sigma} - c_{\vec{k}
\sigma})
\end{array}\right],
\end{equation}
where $\varphi_{\vec{k}} = \arg(\gamma_{\vec{k}})$.

In this basis, where $c_{\vec{k} \sigma}$ and $d_{\vec{k} \sigma}$
represent the electron states in the upper and lower band, respectively,
the Hamiltonian transforms into
\begin{eqnarray}
&&H = \nonumber \\
&& \sum_{\vec{k}} \left\{ \sum_{\sigma}(\ttilde \epsilon_{\vec{k}} - \mu)
c_{\vec{k} \sigma}^{\dagger} c_{\vec{k} \sigma} + \sum_{\sigma}(- \ttilde
\epsilon_{\vec{k}} - \mu) d_{\vec{k} \sigma}^{\dagger} d_{\vec{k}
\sigma} \right. \nonumber \\
&&+ \sqrt{2} J \left[ \sum_{\vec{u}} \Delta_{\vec{u}}
\cos(\vec{k}\cdot \vec{u} - 2 \varphi_{\vec{k}}) (d_{\vec{k}
\uparrow}^{\dagger} d_{-\vec{k} \downarrow}^{\dagger} - c_{\vec{k}
\uparrow}^{\dagger} c_{-\vec{k} \downarrow}^{\dagger}) \right. \nonumber  \\
&& \left.\left. +\sum_{\vec{u}} i \Delta_{\vec{u}}
\sin(\vec{k}\cdot \vec{u} - 2 \varphi_{\vec{k}})
 (c_{\vec{k} \uparrow}^{\dagger}
d_{-\vec{k} \downarrow}^{\dagger} - d_{\vec{k} \uparrow}^{\dagger}
c_{-\vec{k} \downarrow}^{\dagger})\right] + \mbox{H.c.} \right\}. \nonumber
\\\label{ham}
\end{eqnarray}
Here $\ttilde \equiv t^2/t_\bot$
and $\epsilon_{\vec{k}} \equiv |\gamma_{\vec{k}}|^2$.
The eigenvalues are given by
\begin{equation}
E_{\vec{k}} = \pm \sqrt{(\ttilde \epsilon_{\vec{k}})^2  + \mu^2 +
2 J^2 \left( |S_{\vec{k}}|^2 + |C_{\vec{k}}|^2 \right) \pm 2
\sqrt{A}},\label{valuesc}
\end{equation}
where $C_{\vec{k}} = \sum_{\vec{u}}
\Delta_{\vec{u}} \cos(\vec{k}\cdot \vec{u} - 2
\varphi_{\vec{k}})$, $S_{\vec{k}} = \sum_{\vec{u}}
\Delta_{\vec{u}} \sin(\vec{k} \cdot\vec{u} - 2
\varphi_{\vec{k}})$ and
\begin{equation}
A = (\mu^2 + 2J^2 |S_{\vec{k}}|^2) \ttilde^2 \epsilon_{\vec{k}}^2
+ 4J^4 (\mathrm{Re} C_{\vec{k}} \mathrm{Im} S_{\vec{k}} - \mathrm{Im} C_{\vec{k}}
\mathrm{Re} S_{\vec{k}})^2. \label{A}
\end{equation}
If all $\Delta_{\vec{u}}$ are purely real,
i.e. there is no time-reversal symmetry breaking,
then the second term in $A$ is zero and the expression for the dispersion simplifies to
\begin{equation}
E_{\vec{k}} = \pm \sqrt{\left(\ttilde \epsilon_{\vec{k}} \pm \sqrt{\mu^2 +
2J^2S_{\vec{k}}^2 }\right)^2 + 2J^2 C_{\vec{k}}^2 }.
\label{values}
\end{equation}
In this case $S_{\vec{k}}$ only renormalizes the chemical
potential, whereas $C_{\vec{k}}$ plays the main role in the
description of the superconducting order parameter. A comparison
between the Bogoliubov energy dispersion in Eq. (\ref{values}) and
the usual BCS expression shows that $C_{\vec{k}}$ can be
identified with the gap function. However, this name may be
misleading because $C_{\vec{k}}$ does not describe the gap, as in
the example in Eq.~(\ref{s_wave_E}) below.

\begin{figure}
\includegraphics[width = 9cm]{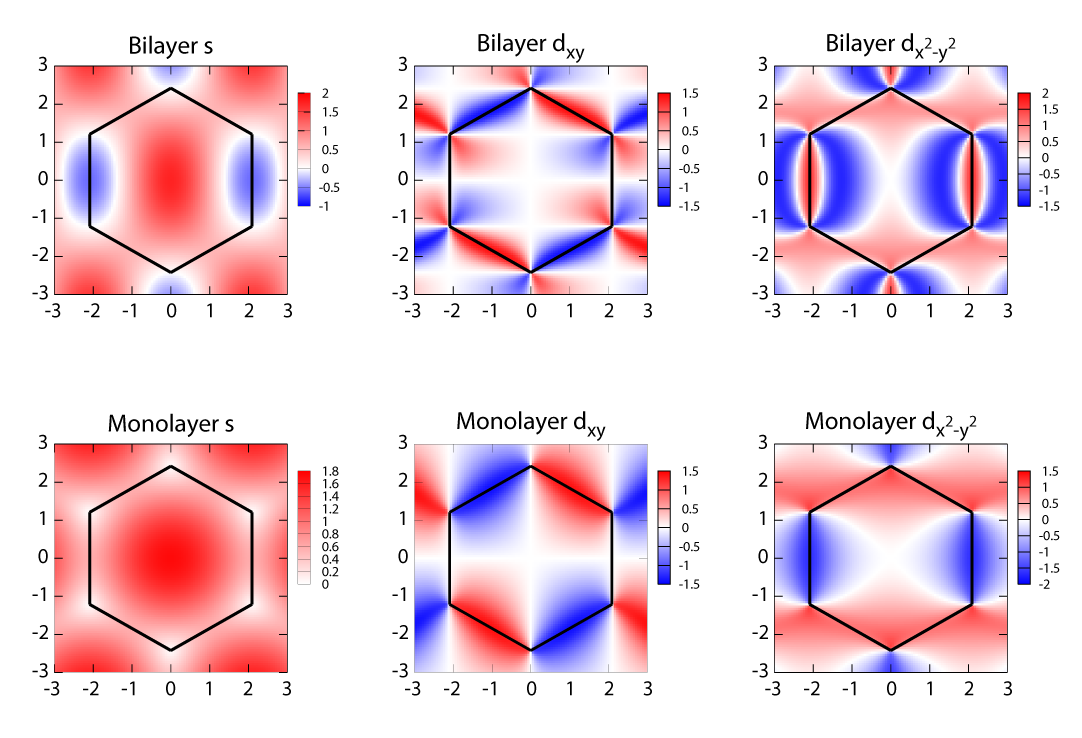}
\caption{(Color online) $C_k$ in the first Brillouin zone
calculated for three possible symmetries on monolayer and
projected bilayer lattices. } \label{fig:04}
\end{figure}

The symmetry analysis of the
order parameter on a honeycomb lattice,\cite{Black} yields the basis vectors which
correspond to
$s$, $d_{x^2-y^2}$ and $d_{xy}$ waves, respectively:
\begin{eqnarray}
\bold\Delta = \left\{
\begin{array}{ccc}
     \Delta \; (1,&1,&1)   \\
     \Delta \; (2,&-1,&-1) \\
     \Delta \; (0,&1,&-1)
\end{array} \right. .
\label{possibilities}
\end{eqnarray}
The gap function $C_{\vec{k}}$ corresponding
to these symmetries is shown in Fig.~\ref{fig:04}, in comparison with the monolayer case. The last two possibilities
belong to a two-dimensional subspace of irreducible representation
of permutation group ${\cal{S}}_3$.\cite{Polleti}
This means that any superposition of these two order parameters, which we may identify with the $d_{x^2-y^2}$ [$(2,-1,-1)$ of Eq.
(\ref{possibilities}) and permutations] and $d_{xy}$ [$(0,1,-1)$ of Eq. (\ref{possibilities}) and permutations] solutions of $d$-wave
superconductivity, is possible from a symmetry point of view. In spite of
this principle possibility, the precise realization of a particular order parameter is a question of energy calculations. One notices that the spatial
point symmetry of the underlying honeycomb lattice is $C_{3v}$, which includes $2\pi/3$ rotations, whereas a transformation from $d_{x^2-y^2}$ to
$d_{xy}$ involves $\pi/4$ rotations. The order parameters thus have a different symmetry than
the underlying lattice, as one may also see in Fig..~\ref{fig:04},
such that the two order parameters do not represent degenerate ground states. Indeed we find, within the BCS mean-field theory, that the $d_{x^2-y^2}$
solution has a lower energy than the $d_{xy}$ solution.

This finding needs to be contrasted to the case of $p$-wave superconductivity on the square lattice.\cite{cg}
In the latter case,
superpositions of the $p_x$ and $p_y$ solutions are also permitted by the symmetry of the order parameter, but both solutions are related to each other
by $\pi/2$ rotations that respect the point symmetry of the underlying (square) lattice. The $p_x$ and $p_y$ solutions are therefore degenerate.

The above arguments indicate that the $C_{3v}$ symmetry of the honeycomb lattice is dynamically broken, only through interactions, via the formation
of a $d_{x^2-y^2}$ order parameter. This is similar to the findings of Poletti
\textit{et al.} in the context of superfluidity of spinless fermions with nearest-neighbor attraction.\cite{Polleti} Also in this case, the
$C_{3v}$ symmetry is dynamically broken. Notice finally that in the small-$J$ limit,
i.e. at weak coupling or in the low-energy limit, the BdG system recovers
the symmetry of the $C_{3v}$ group but has also an (emergent) continuous rotational symmetry that will lead to a
$d_{x^2-y^2} \pm \;i \;\sqrt{3}\; d_{xy}$ instability (see Appendix \ref{appA}).

%%%%%%%%%%%%%%%%

In the case of an  $s$-wave order parameter with $\bold{\Delta} = \Delta \;
(1,1,1)$, a small-wave-vector expansion ($|\vec{q}|a\ll 1$) around the $K$-points yields
\begin{equation}
C_{\vec{K}_{\pm} + \vec{q}} \approx \mp
\frac{\sqrt{3}}{2} q_y a \Delta \, ,\qquad
S_{\vec{K}_{\pm} + \vec{q}} \approx +\frac{\sqrt{3}}{2} q_x a \Delta.
\end{equation}
Thus both
couplings are non-zero and no simple effective picture emerges by
looking at the Hamiltonian in Eq.~(\ref{ham}). The lower excitation
energy branch can be approximated in the small-momentum limit as
\begin{eqnarray}
\nonumber
E_{\vec{q}} &\simeq& \sqrt{\mu^2 - 2 \mu \ttilde \epsilon_{\vec{K}_{\pm}+\vec{q}} +
\frac{3 }{2} J^2 (|\vec{q}| a)^2 \Delta^2}\\
&\simeq& \sqrt{\mu^2 - \frac{3}{2}[3\mu \ttilde  - (J\Delta)^2] (|\vec{q}| a)^2}, \label{s_wave_E}
\end{eqnarray}
where we have used $\epsilon_{\vec{K}_{\pm}+\vec{q}}\simeq 9 (|\vec{q}|a)^2/4$.

If the coupling strengths are such that $E_{\vec{q}}$ has a
minimum at $q = 0$,  that is for $(J\Delta)^2>3\mu\ttilde$, a special
superconducting instability may be realized (if other possibilities,
order parameters, have higher free energy).\cite{Note1}
In the absence of trigonal warping at very low doping, we
obtain a time-reversal invariant superconducting instability
with two kinds of Cooper pairs with $p_x + i p_y$ and $p_x - i
p_y$ pairings.
 Due to the forms of $C_{\vec{k}}$ and $S_{\vec{k}}$ in
the above Hamiltonian in the small momentum limit, $p$-wave Cooper
pairings are expected. For a sufficiently large chemical potential,
one can neglect $S_{\vec{k}}$ in Eq. (\ref{values}) and the
system may be unstable towards a $p_y$ gapless superconductor,
with gap minima on the Fermi surface, i.e. on a circle.

For $\bold{\Delta} = \Delta (2,-1,-1)$,
the small-momentum expansion around the $K$-points yields
\begin{eqnarray}
\nonumber
C_{\vec{K}_{\pm} + \vec{q}}(d_{x^2-y^2}) &\approx& -3 \frac{(q_x^2 -
q_y^2)}{|\vec{q}|^2} \Delta\, , \\
S_{\vec{K}_{\pm} + \vec{q}}(d_{x^2-y^2}) &\approx& \mp 6 \frac{q_x q_y}{|\vec{q}|^2} \Delta
\label{eq:dx2y2}
\end{eqnarray}
and for $\bold{\Delta} = \Delta (0,1,-1)$
\begin{eqnarray}
\nonumber
C_{\vec{K}_{\pm} + \vec{q}}(d_{xy}) &\approx& 2 \sqrt{3} \frac{q_x
q_y}{|\vec{q}|^2} \Delta \, ,\\
S_{\vec{K}_{\pm} + \vec{q}}(d_{xy}) &\approx&
\mp \sqrt{3} \frac{(q_x^2 - q_y^2)}{|\vec{q}|^2} \Delta.
\label{eq:dxy}
\end{eqnarray}
The gap function $C_{\vec{k}}$ thus clearly shows the $d_{x^2-y^2}$ and the $d_{xy}$ symmetry in Eq. (\ref{eq:dx2y2}) and (\ref{eq:dxy}),
respectively.

Notice that one may superpose two waves in the manner
\begin{equation}
C_{\vec{k}}(d\pm id)=C_{\vec{k}}(d_{x^2-y^2}) \pm i \sqrt{3} C_{\vec{k}}(d_{xy}),
\end{equation}
and
\begin{equation}
S_{\vec{k}}(d\pm id)=S_{\vec{k}}(d_{x^2-y^2}) \pm i \sqrt{3} S_{\vec{k}}(d_{xy}) ,
\end{equation}
which is identified with the $d + i d$-wave superconducting phase in the following.
In the small-wave-vector limit, the combined forms of $C_{\vec{k}}$,
\begin{equation}
C_{\vec{K}_{\pm} + \vec{q}} (d+id)\approx \mp i S_{\vec{K}_{\pm} + \vec{q}} \approx 3 (q_x + i
q_y)^2/|\vec{q}|^2 \label{ex_plus}
\end{equation}
and
\begin{equation}
C_{\vec{K}_{\pm} + \vec{q}} (d-id)\approx \pm i
S_{\vec{K}_{\pm} + \vec{q}} \approx 3 (q_x - i q_y)^2/|\vec{q}|^2,
\label{ex_minus}
\end{equation}
restore the rotational symmetry -- they are indeed eigenstates of
rotation in two dimensions with the value of angular momentum
equal to two. Thus a fixed complex combination in real space, either
$d_{x^2-y^2} + i \sqrt{3}\; d_{xy}$ or $d_{x^2-y^2} - i \sqrt{3}\; d_{xy}$,
leads to the same  form of the expansion in small momenta at both valley points,
either (\ref{ex_plus}) or (\ref{ex_minus}).
Because it is the same irrespective of the valley $K$ or $K'$ one
obtains a solution that spontaneously breaks time-reversal
symmetry. Thus we can identify the solution with the broken
time-reversal symmetry $d + i d$ state.
Something similar happens
in the monolayer case, but the $d$-wave symmetry is recognized as
a global dependence of the order parameter on the $\vec{k}$ vector
in the Brillouin zone around the  central $\Gamma$-point (see Ref.
\onlinecite{Linder}) and $p$-wave behavior around $\vec{K}_{\pm}$
points.\cite{uchoa} In the bilayer case the time-reversal symmetry breaking
$d$-wave  order parameter emerges as a property of the low-energy
small-momentum effective description around the $K$ points, as
shown above.

%%%%%%%%%%%%%%%%%%%%%%%%%%%%%%%%%%%%%%%%%%%%%%%%%%%%%%%%%%%%%%%%%%%%%%%

\section {Phase Diagram}

We have found the ground state of our model Hamiltonian for a broad range of $J$ and $\mu$ by minimizing the free energy.
At zero temperature, as a function of the order parameter, it is given by
\begin{equation}
 F = -\sum_{{\vec k}\in \mathrm{IBZ}} \sum_{\alpha=\pm 1}
 E_{{\vec k},\alpha} + 2NJ\sum_{{\vec u}}|\Delta_{\vec u}|^2,
\end{equation}
where the first sum is over all wave vectors $\vec{k}$ in the
first Brillouin zone and two Bogoliubov bands with positive
energies. The ground state is defined as a global minimum of the
free energy in the order parameter space. In the present study, we
concentrate on superconducting order parameters in a variational
approach, and thus we cannot exclude that other correlated
(non-superconducting) phases may have an even lower energy. In the
mean-field approach, superconducting ground states are expected
even for infinitesimal positive values of $J$.

\begin{figure}
\includegraphics[width = 9cm]{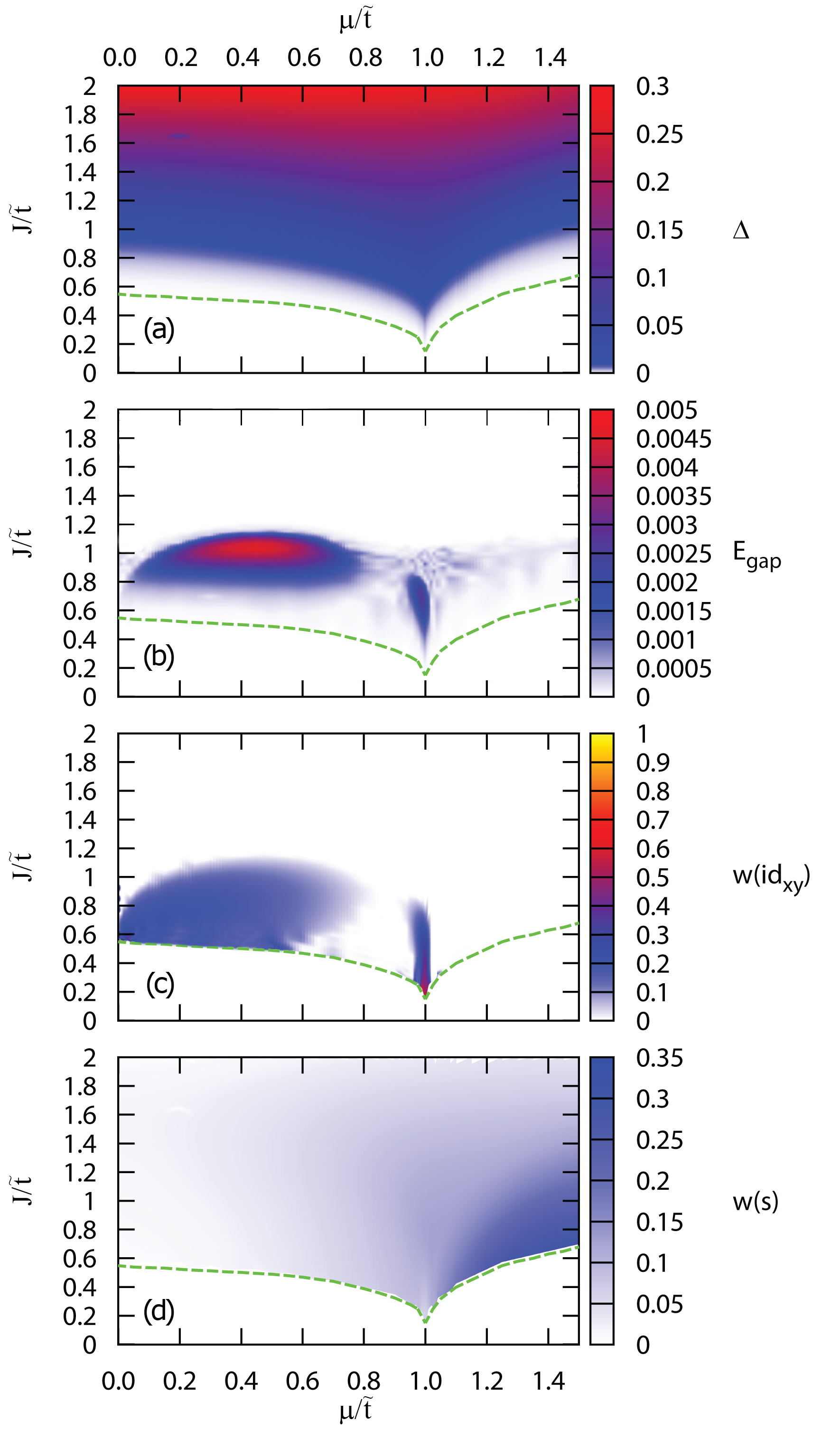}
\caption{(Color online) (a) The order parameter amplitude,
$\Delta$, in the $(\mu,J)$ parameter space, obtained by a
minimization of the free energy, (b) the single-particle
excitation gap, (c) the contribution of $id_{xy}$ and (d) $s$-wave
component in the ground state order parameter. The green dashed
line marks where $\Delta$ drops below $10^{-4}$. Below this line,
our numerics is not reliable. We use $\tilde t = t^2/t_{\bot}$ for
the unit of energy.} \label{fig:05}
\end{figure}

The order parameter space is 6-dimensional, because it is defined
by 3 complex numbers. However, adding the same phase to all three
complex parameters does not modify the physical state, so one can
always make one of the parameters purely real (we set
$\Delta_{d_{x^2-y^2}}$ real) and reduce the order parameter space
dimensionality to 5. We used the amoeba numerical
method\cite{Numrec} to directly minimize the free energy.
Five-dimensional minimization often reveals more than one local
minimum, but we were always able to identify the lowest-lying
state to a satisfying level of certainty. However, for small
values of $J$, the local free-energy minima are extremely shallow,
with energies only slightly lower than the free energy of the
normal state. Such features in the free-energy landscape are
completely clouded by numerical noise due to the discretization of
the first Brillouin zone. Our numerical calculations are therefore
limited to higher values of $J$, which give a solution with the
amplitude of the order parameter larger than $10^{-4}$. This is
marked by the dashed lines in Fig.~\ref{fig:05}.
%, and a short
%discussion of this line in the framework of a weak-coupling
%analysis may be found below [Eq. (\ref{eq:gap})].

Our results are shown on Fig.~\ref{fig:05} where the relevant
quantities are represented by color in the $(\mu,J)$ plane. The
amplitude of the order parameter is shown in Fig.~5(a). Upon small
to moderate doping, the SC instability increases and becomes
particularly favorable at the filling $5/8$, which corresponds to
the chemical potential $\mu / \tilde t =1$, and the van-Hove
singularity in the non-interacting DOS. For further doping the SC
instability decreases. This gives to Fig.~\ref{fig:05}(a) roughly
the look  of the inverse DOS of Fig.~\ref{fig:02}(b). The gap in
the single-particle excitations is shown in Fig.~\ref{fig:05}(b).
It is particularly pronounced in the case of strong mixing of
$d_{x^2-y^2}$ and $i d_{xy}$ symmetry components, as we can see
from Fig.~\ref{fig:05}(c). The contribution of different pairing
symmetries is defined by the ratio $w$ of different components of
$\Delta$, where
\begin{eqnarray}
{\bf \Delta} &=& \Delta_s \hat{e}_s + i \Delta_{is} \hat{e}_s +
\Delta_{d_{xy}} \hat{e}_{d_{xy}} + i \Delta_{id_{xy}}
\hat{e}_{d_{xy}} \nonumber \\
&+& \Delta_{d_{x^2-y^2}} \hat{e}_{d_{x^2-y^2}},
\end{eqnarray}
with $\hat{e}_s = (1,1,1)/\sqrt{3}$, $\hat{e}_{d_{xy}} =
(0,1,-1)/\sqrt{2}$, and $\hat{e}_{d_{x^2-y^2}} =
(2,-1,-1)/\sqrt{6}$.
Fig.~\ref{fig:05}(c) shows the ratio $w(id_{xy}) = |
\Delta_{id_{xy}} | / |\Delta|$, and Fig.~\ref{fig:05}(d) the ratio
$w(s) = | \Delta_{s} | / |\Delta|$. The contributions of $is$ and
$d_{xy}$ components are negligible in all cases, and $d_{x^2-y^2}$
is the dominant component.

\begin{figure}
\includegraphics[width = 9cm]{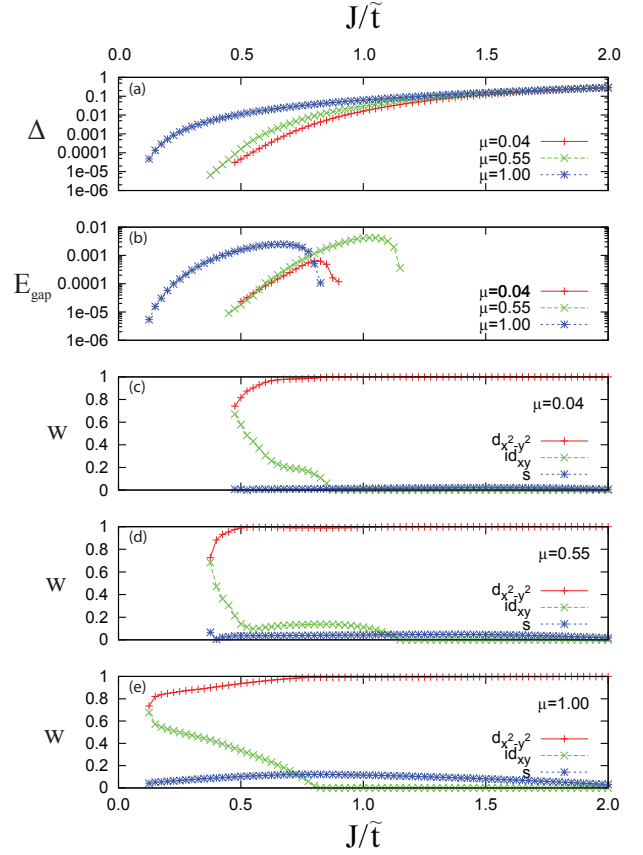}
\caption{(Color online) (a) The order parameter amplitude $\Delta$
and (b) the single-particle excitation gap as a function of $J$,
for $\mu=0.04,0.55,1$. (c)-(e) The contributions of 3 relevant
symmetry components. $d_{x^2-y^2}$ component is the dominant one
for large $J$. The contribution of $id_{xy}$ increases with
decreasing $J$ until the two contributions are equal and we find a
pure $d+id$-wave symmetry. We use $\tilde t = t^2/t_{\bot}$ for
the unit of energy. The data are plotted only above the value for
the coupling $J$ which is numerically significant, as mentioned in the
text (see also the dashed green line in Fig. \ref{fig:05}).
} \label{fig:06}
\end{figure}

The numerical results are, for clarity, also shown on
Fig.~\ref{fig:06} for three chosen values of the chemical
potential, $\mu/\ttilde =0.04, 0.55, 1$. Fig.~\ref{fig:06}(a)
shows a sudden increase in the pairing amplitude with the
increasing interaction $J$ (note the logarithmic scale on the
$y$-axis). For small $J$, the pairing amplitude is much larger for
$\mu/\ttilde=1 $, i.e. at the van-Hove singularity, and in this
case the single-particle excitation gap is also larger due the
strong mixing of $d_{x^2-y^2}$ and $i d_{xy}$ symmetries.
Contributions of relevant components are compared in
Figs.~\ref{fig:06}(c)-(e). At higher values of $J$ one has a pure
$d_{x^2-y^2}$ symmetry, whereas a mixture of $d_{x^2-y^2}$ and
$id_{xy}$ symmetries is found at lower values of $J$. The
contribution of $id_{xy}$ symmetry increases with decreasing $J$
and almost pure $d+id$ symmetries are usually found at the lowest
accessible values of $J$.

Our numerical calculations were performed on processors with 8GB
of RAM which limited the number of $\vec{k}$-points in the first
Brillouin zone to $4000\times 4000$, but we checked that results
do not differ qualitatively even with a much sparser $2000\times
2000$ $\vec{k}$-grid. A much denser and probably a non-uniform
discretization of the first Brillouin zone would be needed to
probe the weak-coupling behavior of our model, that is for values
of $J$ below the dashed lines in Fig. \ref{fig:05}. Notice,
however, that the system in the small-$J$ limit may be treated
analytically within the weak-coupling limit the results of which
are presented in Appendices A and B, for the cases of finite and
zero chemical potential, respectively.

In this weak-coupling regime and at finite chemical potential, we
find that the $d+id$ superconducting order parameter yields the
lowest mean-field energy, when compared to order parameters that
respect time-reversal symmetry (Appendix A),  in agreement with
our numerical results for larger values of $J$. In the weak-coupling limit, in the symmetry-protected subspace of $d_{x^2-y^2}$ and $d_{xy}$
order parameters the complex combination $d_{x^2-y^2}+ i \sqrt{3} d_{xy}$ leads to fully gapped system with no nodes at the Fermi
surface. This means that the gap is proportional to $|C_{\vec{k}}| = const$, and maximum gain in the energy for this
superconducting instability is obtained. Notice that this topological instability
is in line with a theorem for the BCS description, according to which a time-reversal symmetry broken 2D superconducting state has a lower free energy,
as compared to time-reversal symmetric ones, when confronted with two-dimensional representations of the superconducting order parameter.\cite{cg}
Indeed, as mentioned after Eq. (\ref{possibilities}), the $d_{x^2-y^2}$ and $d_{xy}$ components of the order parameter $\bf{\Delta}$ form a
two-dimensional irreducible representation of the symmetry group of the honeycomb lattice. Although the theorem of Ref.~\onlinecite{cg} was
derived for a single band, it is expected also to apply to the present case at finite doping when the higher Bogoliubov band is irrelevant for the
superconducting instability. This instability occurs at any
strength of attractive interaction at finite doping since the gap
opens as
\begin{equation}\label{eq:gap}
J\Delta \propto
\exp\left[-\frac{8\pi}{\sqrt{3}}\frac{1}{\rho(\mu)J}\right]\end{equation}
(see Appendix A), in terms of the DOS $\rho(E_F)$ at the Fermi
level $E_F$. This is simply the BCS expression with the pairing potential equal to
$J$.

Finally, we notice that the weak-coupling analysis yields a
different picture at zero-doping (Appendix B), where a
time-reversal-symmetric superconducting order parameter (with any
real  combination of $d_{x^2-y^2}$ and $d_{xy}$) is
energetically favored.

\section{Possible relevance for bilayer graphene}

In the following we will discuss possible relevance of our model for the
physics of bilayer graphene. With an estimate \cite{Castro,Wehling} for the Coulomb on-site repulsion, $U \sim 10$ eV, intralayer
nearest-neighbor hopping,\cite{data}
$t \sim 3$ eV, and interlayer hopping,\cite{data} $t_\bot  \sim 0.4$ eV, bilayer graphene  may have a tendency to develop
strongly-correlated electron phases. Notice that, although similar energy scales are found in monolayer graphene, the latter is to
great accuracy described in terms of (quasi-)free electrons because of a vanishing DOS at the Fermi level, in
the
absence of intensive doping.\cite{Castro,Kotov,Goerbig} On the contrary, electronic correlations are much more efficient in
bilayer graphene as a consequence of the finite DOS even at the band-contact points. This finite DOS
may also be invoked when considering screening. Whereas screening is highly inefficient in monolayer graphene, and one needs then to
take into account the long-range nature of the electronic interaction potential, the screening properties in bilayer graphene are
similar to those in usual 2D electron systems with a parabolic band dispersion, albeit with a rather small band mass
($\sim 0.05m_0$, in terms of the bare electron mass). In this sense, an approach based on the Hubbard model, as used here excluding nearest and
further-neighbor
interactions, is better justified in bilayer than in monolayer graphene.
However, this remains a strong approximation, as in the case of 2D electrons in GaAs heterostructures, and numerical calculations
indicate that longer-range terms remain relevant also in bilayer graphene.\cite{Wehling}

Generally, the interplay between a strong on-site repulsion $U$ and the hopping terms $t$ and $t_{\perp}$ leads to antiferromagnetic
Heisenberg-type exchange interactions, $J \sim t^2/U \sim 1\,\text{eV}$ between nearest neighbors in the same layer and
$J_\bot \sim t_\bot^2/U \sim 16\,\text{meV}$ between nearest neighbors in opposite layers. Although clear evidence for antiferromagnetism
is lacking in bilayer graphene, the quadratic dispersion of juxtaposed conduction and valence bands (together with the non-zero density
of states) favor antiferromagnetic fluctuations.\cite{MacDonald} Because the low-energy electrons move preferentially on the B1 and
B2 sublattice sites, one needs to estimate an effective exchange interaction between them that may be obtained from a perturbative expansion,
$J_\text{eff} \sim J^2 J_\bot/t_\bot^2 \sim t^4/U^3 \sim 100\,\text{meV}$.

Remember that the effective hopping parameter in the projected honeycomb lattice (between the B1 and B2 sites)
is a more subtle issue because it is derived in the limit where $t_\bot\gg t$, in contrast to the natural order in bilayer graphene. In order to
make a comparison between our effective model and that of bilayer graphene, in view of the correlated phases we consider, it is therefore more appropriate
to define the effective hopping indirectly from the value of $J_\text{eff}$ and $U$,
$J_\text{eff} \sim   t_\text{eff}^2/U$, which yields a value of $t_\text{eff} \sim 1\,\text{eV}$ that should replace the value $\tilde{t}$ in
the previous sections.

Therefore modeled with two effective parameters, $J_\text{eff}$
and $t_\text{eff}$, bilayer graphene may be compared with the
effective honeycomb lattice considered in our paper and the
corresponding $t-J$ model. The main feature of bilayer graphene
appears to be that  $J_\text{eff}\sim 0.1 t_\text{eff} \ll
t_\text{eff}$ and in considering the relevance of our model we
should confine ourselves to weak couplings, and small or moderate
dopings; because we simplified the high-momentum physics of the
bilayer (by considering the large $t_\bot$ limit) we should
confine ourselves to lower dopings. First one sees from Fig.
\ref{fig:05} that the gaps are in the meV range (2 to 5 meV for
the maximal gaps) if one considers the energy scale
$t_\text{eff}\sim J\sim 1$ eV. Thus our results indicate very
small energy scales that are unlikely to be resolved in today's
graphene samples. Furthermore we should use $t_\text{eff}$ and
$J_\text{eff}$ for $t$ and $J$  for the exponent in the
weak-coupling analysis in the Appendix A. Because we estimate $
t_\text{eff}/J_\text{eff} \sim 10$, the weak-coupling analysis
yields an exponential suppression  and gaps below 1 meV, in
agreement with our numerical findings shown in Fig. \ref{fig:05}.

\section{Conclusions}

We presented an analysis of a model of honeycomb bilayer with attractive interactions that
(1) supports $d + i\; d$ superconductivity with the canonical effective (low-momentum) description
$\sim (k_x + i k_y)^2$ at both valley points, and (2) at moderate and strong couplings transforms into $d_{x^2-y^2}$ superconductivity.
The implied $tJ$ model may be relevant for future investigations of such a complex and intriguing   system as the graphene bilayer. We discussed the
possibility of a superconducting
instability in this framework and concluded  that $d + i d$ is the leading
superconducting instability in the case of the graphene bilayer at moderate dopings and low energy scales.

We would like to point out also to the difference between
monolayer and bilayer case that follows form the symmetry analysis
of the simple model with attractive interactions and ensuing short
range order parameter on both lattices. In the effective description
around $\vec{K}$ points $s$-wave and $p$-wave are found
\cite{Black,Linder} in the monolayer case, and $p$-wave and $d$-wave
in the bilayer case. The bilayer honeycomb lattice appears at moderate
dopings as yet another stage on which time reversal symmetry breaking $d$-wave
superconductivity may appear (see \cite{Baskaran,Black,Honerkamp,Pathak,Pellegrino,Gu} for moderately doped monolayer) and may be driven by similar
physics as in the case of predicted instabilities at special (very high) dopings of honeycomb monolayer \cite{Ch,Ab}. In the case we presented the canonical \cite{rg} low momentum description, $\sim (k_x + i k_y)^2$, holds due to the quadratically dispersing Dirac electrons.

\begin{acknowledgments}
We thank A.M. Black-Schaffer, M. Civelli, M. Franz, and
 Y. Hatsugai for useful discussions. Furthermore, we thank D. Tanaskovi\'c for support and his implication at the early stage of this project. J.V. and M.V.M.
are supported by the Serbian Ministry of Education and Science under
project No. ON171017, and M.O.G. by the ANR (Agence Nationale de la Recherche) project NANOSIM GRAPHENE under Grant No.
ANR-09-NANO-016. The authors acknowledge financial support
from bilateral MES-CNRS 2011/12 program. This research was funded in part by the National Science
Foundation under Grant No. NSF PHY05-51164; M.V.M. and M.O.G. acknowledge the hospitality of KITP, Santa Barbara.
Numerical simulations were run on the AEGIS e-Infrastructure, supported in
part by FP7 projects EGI-InSPIRE, PRACE-1IP and HP-SEE.
\end{acknowledgments}

\appendix

\section{Weak-coupling analytical solution at finite chemical potential}
\label{appA}

Here, we present briefly the weak-coupling analysis of
superconducting order in the effective bilayer model. In order to
simplify the notation, we use the letter $t$  to denote the
effective hopping $\tilde{t}$. The DOS at the Fermi level,
$\rho(E_F)$, is on the order of the inverse hopping parameter
$1/t$. Notice that, if only a parabolic band is taken into account
it remains fixed at its $E_F=0$ value, but corrections to the
parabolic approximation immediately yield a contribution that
varies linearly with the Fermi level, in agreement with the DOS
plotted in Fig. \ref{fig:02}(b).

In the case when $\bold{\Delta} = \Delta (1,1,1)$, a weak-coupling BCS analysis
that takes into account only electrons in the lower Bogoliubov band gives
\begin{equation}
J \Delta =  \sqrt{2 t E_c} \exp\left(- 24 \sqrt{3} \pi
\frac{t}{\mu \rho(E_F) J}\right),
\end{equation}
with   $E_c$ as an energy cut-off around the Fermi value, for the
solution, and
\begin{equation}
\frac{\delta E_{MF}^{p}}{N} = - (J \Delta)^2
\frac{\mu \rho(E_F)}{t} \frac{1}{4 \sqrt{3} \pi},
\end{equation}
for the gain in the mean-field energy, $ \delta E_{MF}$, by the
pairing instability.

 The weak coupling BCS analysis in the case of electron doping $(\mu  > 0)$ for $d_{x^2 - y^2}$ and
$d_{x^2 - y^2} + i \sqrt{3}\;d_{xy}$ gives
\begin{equation}
J \Delta_d = \frac{\sqrt{2}}{3}  E_c \exp\left(-\frac{8
\pi}{\sqrt{3}} \frac{1}{\rho(E_F) J} + \frac{1}{2}\right),\label{a1}
\end{equation}
 for the solution which we denoted by $\Delta = \Delta_d$, and
 \begin{equation}
J \Delta = \sqrt{\frac{2}{3}}  E_c \exp\left(- \frac{8
\pi}{\sqrt{3}} \frac{1}{\rho(E_F) J} + \frac{1}{2}\right),\label{a2}
\end{equation}
in the case of $d_{xy}$ wave. For the energy gain one obtains
\begin{equation}
\frac{\delta E_{MF}(d_{x^2 - y^2})}{N} =  \frac{\delta
E_{MF}(d_{xy})}{N} = - (J \Delta_d)^2 \rho(E_F) \frac{3
\sqrt{3}}{4 \pi},
\end{equation}
and for a $d_{x^2 - y^2} + i \sqrt{3}\; d_{xy}$ wave, one finds
\begin{equation}
\frac{\delta E_{MF}^{d}}{N} = - (J \Delta_d)^2 \rho(E_F)
\frac{3 \sqrt{3}}{2 \pi}.
\end{equation}
Because of its twice lower mean-field energy, the $d_{x^2 - y^2} + i \sqrt{3}d_{xy}$ time-reversal symmetry breaking
instability, which we call in short $d$-wave, is more likely than  $d_{x^2-y^2}$ and $d_{xy}$-wave order parameters.
 In the large-doping limit,
the energy minimization is also much more efficient for $d$-wave
than $p_y$-wave  as seen in  the small value of the ratio
\begin{equation}
\frac{\delta E_{MF}^{p}}{\delta E_{MF}^{d}} = \frac{\mu}{2 E_c}
\exp\left[ - \frac{2 \pi \times 8}{\sqrt{3}} \frac{1}{\rho(E_F) J} \left(\frac{9 t}{2
\mu} - 1\right)\right],
\end{equation}
for $ \mu < \frac{9 t}{2}$. The most natural choice for $E_c$ is to
be of the order of $\mu$ as a first energy scale when we start from
the smallest one, i.e. $J$.
The time-reversal symmetry breaking $d$-wave solution
of our BCS mean-field Hamiltonian is also expected from a theorem
proved in Ref. \onlinecite{cg}. The theorem was derived for 2D one-band models that reveal both time-reversal symmetry
and a point symmetry described by the dihedral group $D_n$ [or the O(2) rotation symmetry in the case of continuum models]. It states that generally
a time-reversal symmetry breaking superconducting state has a lower free energy than time-reversal symmetric ones if one is confronted with a
2D representation of the symmetry group. In the case of weak coupling that we consider
here, i.e. $J \ll \mu$, and $\mu > 0$ (electron doping), we have an
effective one-band theory of electrons to which
the theorem can be applied. Also the dispersion of the complex $d$-wave order parameter is more complicated in our case
(than in Ref. \onlinecite{cg}) as can be seen in Eqs.(\ref{valuesc}) and (\ref{A}). But in the weak coupling limit the $J^4$ term
can be neglected in Eq. (\ref{A}), and we obtain expressions that are reminiscent to those of Ref. \onlinecite{cg}.

In the following we investigate  more closely an effective
low-energy description of the $d$-wave instability, in the case of
high electron doping, and discuss only the lower energy Bogoliubov
band. Therefore our effective Hamiltonian is
\begin{equation}
H_e = \sum_{\vec{k} \sigma}(t \epsilon_{\vec{k}} - \mu)
c_{\vec{k}\sigma}^{\dagger} c_{\vec{k}\sigma} + \sum_{\vec{k}}\left(
\Delta_{\vec{k}} c_{\vec{k}\uparrow}^{\dagger}
c_{-\vec{k}\downarrow}^{\dagger} + \text{H.c.}\right)
\end{equation}
where $\Delta_{\vec{k}} \sim (k_x - i k_y)^2/|k|^2$. In the
weak-coupling BCS analysis it can be easily shown that the
Hamiltonian is completely equivalent to the one with
$\Delta_{\vec{k}} \sim (k_x - i k_y)^2$, because both Hamiltonians
have an effective description on a Fermi circle defined by $t
\epsilon_{\vec{k}} = \mu$. With this adjustment we have exactly
the form of the BCS Hamiltonian studied in Ref. \onlinecite{rg} on
time-reversal symmetry breaking superconductors in two dimensions.
In the so-called weak-pairing case for finite $\mu > 0$ that we
want to study, the minimum of Bogoliubov excitations moves to
finite values of $\vec{k}$, $t \epsilon_{\vec{k}} = \mu$, i.e. to
the Fermi surface of free particles. The Cooper pair wave function
$g(\vec{r})$ may be a non-universal function of $|\vec{r}|$ where
$\vec{r}$ is the relative coordinate of the pair. On the other
hand, the dependence of the function on the angle of vector
$\vec{r}$ is fixed and can  easily be derived in the Bogoliubov
formalism to be $g(|\vec{r}|) \propto \frac{\bar{z}}{z}\propto
(x-iy)^2$ where $z = x + i y$ is the two-dimensional complex
coordinate. Thus the relative angular momentum of the Cooper pair
is $l = - 2$. The weak-pairing phase is topological, gapped in the
bulk because $\mu > 0$, and possesses a doublet of spin 1/2 Dirac
edge modes \cite{rg}. In our case, because of the fermion doubling
on the honeycomb lattice and the existence of the two $\vec{K}$
points (valleys) [and because around each one we have the same
effective description given by Hamiltonian in Eq.(\ref{ham})], we
expect four Dirac modes on the edge.

\section{Weak coupling analytical solution at zero chemical potential}
\label{appB}

In the weak coupling limit at $\mu = 0$, when both Bogoliubov bands are
taken into account we find for $d_{x^2-y^2}$ symmetry
\begin{equation}
J \Delta^d =  \frac{E_c}{3} \exp\left(-
\frac{\frac{8}{3  J} - 11 c}{2 c}\right),
\end{equation}
with $c \equiv \frac{1}{2 \pi \sqrt{3}} \frac{1}{t}$, for the solution, and
\begin{equation}
\frac{\delta E_{MF}^{d_{x^2-y^2}}}{N} = - \frac{9}{2} c \; (J \Delta^d)^2
,
\end{equation}
for the energy gain. On the other hand for $d + i d$ symmetry we find
\begin{equation}
J \Delta^{d+id} =  \frac{\sqrt{2} E_c}{3} \exp\left(-
\frac{\frac{8}{3  J} - 5 c}{2 c}\right),
\end{equation}
and
\begin{equation}
\frac{\delta E_{MF}^{d+id}}{N} = - 9 c \; (J \Delta^{d+id})^2
.
\end{equation}
Because
\begin{equation}
\frac{\delta E_{MF}^{d+id}}{\delta E_{MF}^{d_{x^2-y^2}}} =
\frac{\delta E_{MF}^{d+id}}{\delta E_{MF}^{d_{xy}}} = 4 e^{-6},
\end{equation}
any real combination of $d_{x^2-y^2}$ and $d_{xy}$ waves is more likely than $d + i d$ wave.


\begin{thebibliography}{9}

\bibitem{Castro} A.~H. Castro Neto, F. Guinea, N.~M.~R. Peres, K.~S. Novoselov, and A.~K. Geim, Rev. Mod. Phys. {\bf 81}, 109 (2009).
\bibitem{Kotov}V. N. Kotov, B. Uchoa, V. M. Peirera, A. H. Castro Neto, and F. Guinea, Rev. Mod. Phys. {\bf 84}, 1067 (2012).
\bibitem{Goerbig}M. O. Goerbig, Rev. Mod. Phys. {\bf 83}, 1193 (2011).
\bibitem{Martin} J. Martin, B. Feldman, T. Weitz, M. Allen, and A.
Yacoby, Phys. Rev. Lett. {\bf 105}, 256806 (2010).
\bibitem{Weitz} R.~T. Weitz, M.~T. Allen, B.~E. Feldman, J. Martin, and A.
Yacoby, Science {\bf 330}, 812 (2010).
\bibitem{Freitag} F. Freitag, J. Trbovic, M. Weiss, and C.
Schonenberger, Phys. Rev. Lett. {\bf 108}, 076602 (2012).
\bibitem{Lau} J. Velasco Jr., L. Jing, W. Bao, Y. Lee, P. Kratz, V. Aji, M. Bockrath, C.~N. Lau, C. Varma, R. Stillwell, D. Smirnov, F. Zhang, J. Jung, A.H. MacDonald, Nature Nanotech. {\bf 7}, 156 (2012).
\bibitem{Qi} X.-L. Qi and S.-C. Zhang, Rev. Mod. Phys. {\bf 83}, 1057 (2011).
\bibitem{Min} H. Min, G. Borghi, M. Polini, and A.~H. MacDonald, Phys.
Rev. B {\bf 77}, 041407(R) (2008).
\bibitem{Nandkishore} R. Nandkishore and L. Levitov, Phys. Rev. Lett
{\bf 104}, 156803 (2010); Phys. Rev. B {\bf 82}, 115124 (2010).
\bibitem{Zhang} F. Zhang, H. Min, M. Polini, and A.~H. MacDonald,
Phys. Rev. B {\bf 81}, 041402(R) (2010).
\bibitem{Vafek} O. Vafek and K. Yang, Phys. Rev. B {\bf 81},
041401(R) (2010).
\bibitem{FZhang} F. Zhang, J. Jung, G.~A. Fiete, Q. Niu, and A.~H. MacDonald, Phys. Rev. Lett. {\bf 106}, 156801 (2011).
\bibitem{Jung} J. Jung, F. Zhang, and A.~H.  MacDonald, Phys. Rev. B {\bf 83}, 115408 (2011).
\bibitem{Lemonik} Y. Lemonik, I.~L. Aleiner, C. Toke, and V.~I. Falko, Phys. Rev. B {\bf 82}, 201408 (2010).
\bibitem{ZhangMacD} F. Zhang and A.~H. MacDonald, Phys. Rev. Lett. {\bf 108}, 186804 (2012).
\bibitem{Vafek_three} R.~E. Throckmorton and  O. Vafek, Phys. Rev. B {\bf 86}, 115447 (2012). 
\bibitem{Kharitonov} M. Kharitonov, arXiv:1109.1553.
\bibitem{Scherer} M.~M. Scherer, S. Uebelacker, and C. Honercamp, Phys. Rev. B {\bf 85}, 235408 (2012). 
\bibitem{McFa} E. McCann and V.~I. Falko, Phys. Rev. Lett. {\bf 96}, 086805 (2006).
\bibitem{Vafek_two} O. Vafek, Phys. Rev. B {\bf 82}, 205106 (2010).
\bibitem{Black} A.~M.  Black-Schaffer and S. Doniach, Phys. Rev. B
{\bf 75}, 134512 (2007).
\bibitem{Bena}C. Bena and L. Simon, Phys. Rev. B {\bf 83}, 115404 (2011).
\bibitem{Polleti} D. Poletti, C. Miniatura, B. Gremaud, Euro. Phys.
Lett. {\bf 93}, 37008 (2011).
\bibitem{Note1} Notice that other symmetries
of the superconducting order parameter are indeed energetically favored at higher doping, as
is shown in the weak-coupling analysis presented in Appendix A.
\bibitem{Linder} J. Linder, A.~M. Black-Schaffer, T. Yokoyama, S.
Doniach, and A. Sudbo, Phys. Rev. B {\bf 80}, 094522 (2009).
\bibitem{uchoa}B. Uchoa and A. H. Castro Neto, Phys. Rev. Lett. {\bf 98}, 146801 (2007).
\bibitem{Numrec} {\textit {Numerical Recipes: The Art of Scientific
Computing}} (Cambridge University Press, Cambridge, 2007).
\bibitem{Wehling} T.~O. Wehling, E. Sasioglu, C. Friedrich, A.~I. Lichtenstein,
M.~I. Katsnelson, and S. Blugel, Phys. Rev. Lett. {\bf 106}, 236805 (2011).
\bibitem{data} K. Zou, X. Hong, and J. Zhu, Phys. Rev.B {\bf 84}, 085408 (2011)  and
references therein.
\bibitem{cg} M. Cheng, K. Sun, V. Galitski, and S. Das Sarma,
Phys. Rev. B {\bf 81}, 024504 (2010).
\bibitem{MacDonald} A.~H. MacDonald, J. Jung, and F. Zhang, Phys. Scr. {\bf 146}, 014012 (2012). 
\bibitem{Baskaran} G. Baskaran, Phys. Rev. B {\bf 65}, 212505 (2002).
\bibitem{Honerkamp} C. Honerkamp, Phys. Rev. Lett. {\bf 100}, 146404 (2008).
\bibitem{Pathak} S. Pathak, V.~B. Shenoy, and G. Baskaran, Phys. Rev. B {\bf 81}, 085431 (2010).
\bibitem{Pellegrino}  F. M. D. Pellegrino, G. G. N. Angilella, and R. Pucci, Eur. Phys. J B 76, 469 (2010).
\bibitem{Gu} Z.-C. Gu, H.-C. Jiang, D.~N. Sheng, H. Yao, L. Balents, and X.-G. Wen, arXiv:1110.1183.
\bibitem{Ch} R. Nandkishore, L. Levitov, and A. Chubukov, Nature Physics {\bf 8}, 158 (2012).
\bibitem{Ab} M. Kiesel, Ch. Platt, W. Hanke, D.A. Abanin, R. Thomale, Phys. Rev. B {
bf 86}, 020507(R) (2012). 
\bibitem{rg} N. Read and D. Green, Phys. Rev. B {\bf 61}, 10267 (2000).
\end{thebibliography}
\end{document}